\documentclass[aps,prx,twocolumn]{revtex4-1}

\usepackage{amsmath}
\usepackage{amsthm}
\usepackage{amsfonts}

\usepackage{bbm,graphicx,bm,hyperref,color}   

\usepackage{multirow}
\usepackage{blkarray, bigstrut}

\def\ave#1{\langle #1 \rangle}
\def\ii{{\rm i}}
\def\sx{\sigma^{\rm x}}
\def\sy{\sigma^{\rm y}}
\def\sz{\sigma^{\rm z}}
\def\tr#1{{\rm tr}(#1)}
\def\1{\mathbbm{1}}

\def\ket#1{{| #1 \rangle}}

\def\braket#1#2{{\langle #1 | #2 \rangle}}

\def\ee#1{{\rm e}^{#1}}
\def\Uhaar{U_{\rm Haar}}
\def\Vhaar{V}

\def\aI{a_{\rm I}}
\def\aII{a_{\rm II}}
\def\bI{b_{\rm I}}
\def\bII{b_{\rm II}}
\def\uI{u_{\rm I}}
\def\uII{u_{\rm II}}
\def\rI{\rho_{\rm I}}
\def\rII{\rho_{\rm II}}
\def\MC{MC }
\def\Rc{\check{R}}
\def\T{\mathbbm{T}(k)}

\def\tit#1{{\em #1},}

\def\PRL{}
\def\app{Appendix}

\begin{document}

\title{
 Integrability is generic in homogeneous U(1)-invariant nearest-neighbor qubit circuits
}

\author{Marko \v Znidari\v c, Urban Duh, and Lenart Zadnik}
\affiliation{Physics Department, Faculty of Mathematics and Physics, University of Ljubljana, 1000 Ljubljana, Slovenia}

\date{\today}

\begin{abstract}
Integrability is an exceptional property believed to hold only for systems with fine-tuned parameters. Contrary, we explicitly show that in homogeneous nearest-neighbor qubit circuits with a U(1) symmetry, i.e., circuits that repeatedly apply the same magnetization-conserving two-qubit gate, this is not the case. There, integrability is generic: all such brickwall qubit circuits are integrable, even with a randomly selected gate. We identify two phases with different conservation laws, transport properties, and strong zero edge modes. Experimentally important is the fact that varying any one of the parameters in the generic U(1) gate, one will typically cross the critical manifold that separates the two phases. Finally, we report on an unconventional time-reversal symmetry causing the system with open boundary conditions to be in the orthogonal class, while the one with periodic boundary conditions is in the unitary class.
\end{abstract}




\maketitle

\section{Introduction} 
Integrable models are indispensable in theoretical physics because they offer a rare glimpse into the full analytical structure of otherwise complex phenomena. Integrability is an exact mathematical concept related to the existence of a sufficient number of conserved quantities. In classical physics a system with $d$ degrees of freedom is said to be integrable if it has (at least) $d$ conserved quantities in involution, a canonical example being the Kepler's two-body problem. Quantum integrability is often described similarly -- one has to have a sufficient number of local conserved operators~\cite{JS}. Finding integrable quantum many-body systems is particularly important due to the exponential complexity of numerical approaches. Two notable examples of many-body integrable systems are the Heisenberg spin-$1/2$ model of quantum magnetism and the 2D Ising model of phase transitions. All these models are solvable by the so-called Yang-Baxter formalism~\cite{baxter}, a definition of integrability which we use herein. A particularly appealing feature of the Yang-Baxter integrability is that the full integrable structure follows from a solution of a Yang-Baxter equation, i.e., from the so-called $R$ matrix [see Eq.~\eqref{eq:YB}]. That structure includes an infinite number of local conserved operators which can be used to analytically study the dynamics, the transport phenomena, etc. -- see Ref.~\cite{jstat} for a collection of reviews.

\begin{figure}
  \centerline{\includegraphics[width=0.49\textwidth]{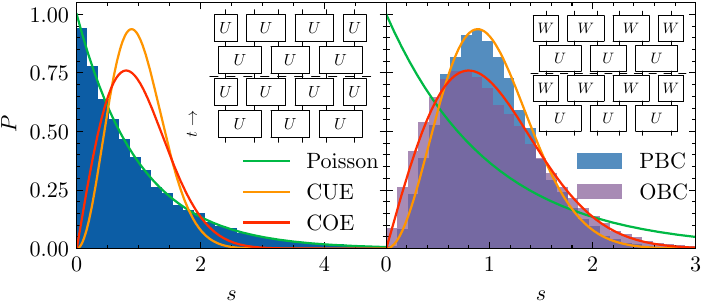}}
  \caption{Level-spacing distribution of eigenphases of a Floquet propagator can often suggest integrability or chaos~\cite{haake,not}. A homogeneous magnetization-conserving qubit circuit with a Haar-random gate and periodic boundary conditions (PBC, left) shows Poissonian statistics, suggesting integrability, while a circuit with two different gates in consecutive layers (right) is chaotic with an additional time-reversal symmetry (see \app~\ref{sec:appA}) in the case with open boundary conditions (OBC). Numerics is done on $L=16$ qubits, resolved over magnetization (OBC \& PBC), momentum (PBC), and space-time symmetry~\cite{spacetime} (homogeneous).}
  \label{fig:lsd}
\end{figure}

With a rapid progress of quantum computers and simulators~\cite{jepsen20,bloch22,morvan22,joshi22}, where the natural setting is that of a circuit made of local gates, having integrable quantum circuits is desirable. Same-gate circuits, in which one repeatedly applies the same unitary gate, are especially appealing, since they are simpler to realize experimentally, and they can be theoretically studied using the Floquet formalism. Inspired by integrable light-cone discretizations of quantum field theories~\cite{destri87,volkov94}, some integrable Floquet systems have been found in the recent decade, for example, a nearest-neighbor circuit with a gate of an XXX or XXZ type~\cite{gritsev,vanicat18,ljubotina19}. Those initial findings have since been extended in several directions~\cite{austen22,gritsev24,hutsalyuk24}, including higher dimensional local space~\cite{ratchet}, unitary gates with a larger range~\cite{sarang18,vernier23,korepin24,balazs21,balazs21b}, as well as non-unitary circuits~\cite{sa21,paletta24}. Nevertheless, integrability is still regarded as a fine-tuned property. For instance, a generic perturbation of any of the above known integrable Floquet systems is believed to break integrability, leading to chaos~\cite{gaspard,Braun}. In this work we show that this is not always the case. Specifically, using the $R$ matrix of the asymmetric six-vertex model~\cite{sogo82,Marius13,vieira18}, we will explicitly show that U(1)-invariant circuit composed of a magnetization-conserving (MC) two-qubit Haar-random gate arranged in a brickwall pattern is integrable (see Fig.~\ref{fig:lsd}). Note that the U(1) symmetry alone does not in any way guarantee integrability. On the contrary, a typical system with magnetization conservation is chaotic, e.g., the 1D XXZ spin chain in a staggered field, or the 2D Hubbard model.

\section{Magnetization-conserving circuits}

The circuits we study consist of two-qubit unitary gates that preserve magnetization, i.e., they have a U(1) symmetry. The gate can be expressed as $U=\ee{-\ii \tau h}$, where
\begin{eqnarray}
  h_{1,2} &=&\sx_1\sx_2+\sy_1\sy_2+\Delta \sz_1 \sz_2+M (\sz_1+\sz_2)\notag\\
  & &+B(\sz_2-\sz_1)+D(\sx_1 \sy_2-\sy_1 \sx_2) + A\,\1
  \label{eq:H}
\end{eqnarray}
acts on two neighboring qubits, labeled by $1$ and $2$. Parameters $A$ and $M$, associated with two trivially conserved operators, i.e., the identity $\1$ and the magnetization $\sz_1+\sz_2$, are irrelevant, as they only determine the phase. One therefore has four relevant parameters: the gate duration $\tau$, the anisotropy $\Delta$, the staggered magnetic field $B$, and the Dzyaloshinskii--Moriya coupling $D$ (current). While we will return to the above parametrization later, for the proof of the circuits' integrability  it will be easier to express the gate using the standard Hurwitz parametrization~\cite{hurwitz} of a Haar-random \MC unitary~\cite{foot0}:
\begin{equation}
\!\!\!\Uhaar\!=\!\left[\!
   \begin{blockarray}{@{\,}cccc@{\,}}
\bigstrut[t]
           \ee{\ii \delta} & 0 & 0 & 0 \\
        \begin{block}{@{}c\BAmulticolumn{2}{!{}c!{}}@{}c}
            0 & \multirow{2}{*}{\raisebox{-1pt}{\framebox(18,18){$\Vhaar$}}} & \,\,0\\
           0 & & \,\,0 \\
        \end{block}
        \BAnoalign{\vskip -5.5ex}
            0 & 0 & 0 & \,\ee{\ii \delta}       
   \end{blockarray}\!\right]\!,\,
   \Vhaar\!=\!\ee{\ii \alpha}\!\begin{bmatrix}
		\sin{\!\varphi}\ee{-\ii \chi} & \!\cos{\!\varphi}\ee{-\ii\vartheta}\\
\cos{\!\varphi}\ee{\ii\vartheta} & \!-\sin{\!\varphi}\ee{\ii \chi}
     \end{bmatrix}\!.   
   \label{eq:Haar}
\end{equation}
Here, we removed the irrelevant phase corresponding to the conserved magnetization (the field $M$), which allows us to write the same phase in the two corners of $\Uhaar$. A generic $\Uhaar$ is obtained by picking the parameters from a distribution that is uniform with respect to an invariant U(1)-preserving measure ${\rm d}\mu \sim {\rm d}(\sin^2{\varphi}) {\rm d}\chi\, {\rm d}\alpha\, {\rm d}\vartheta\,{\rm d}\delta$, with $\alpha,\chi,\vartheta,\delta \in [0,2\pi)$, and $\varphi \in [0,\pi/2)$. Quantum circuits utilizing such \MC random gates have received a lot of attention recently -- see, e.g., Refs.~\cite{pollmann18,vedika18,deluca}. To establish integrability of a circuit we need to find a unitary solution of the Yang-Baxter equation that will be equal to our gate $\Uhaar$, and show that it can span {\em all} \MC Haar-random gates. The latter is not obvious~\cite{foot1}, despite the integrability~\cite{sogo82,Marius13,vieira18} of all translationally invariant Hamiltonians with local densities~\eqref{eq:H}~\cite{foot5}. Once that is done, all nice integrability consequences follow from the Yang-Baxter formalism~\cite{baxter}. The next section recalls the basic formulation of integrable circuits~\cite{vanicat18,ljubotina19}. 
    
\section{Integrable circuits} 

The Yang-Baxter equation in its braid form is,
\begin{equation}
\!\!\!\Rc_{1,2}(x) \Rc_{2,3}(\!x+y\!) \Rc_{1,2}(y)\!=\!\Rc_{2,3}(y) \Rc_{1,2}(\!x+y\!) \Rc_{2,3}(x),
\label{eq:YB}
\end{equation}
where we additionally demand $\Rc(-x)\Rc(x)=\mathbbm{1}$ and $\Rc(0)=\mathbbm{1}$. It is the basic building block of integrable systems and a defining equation of $\Rc$, having many solutions corresponding to different integrable systems. Once a solution $\Rc$ is found  the conserved charges can be constructed with the help of a transfer matrix. While $\Rc$ can depend on many free parameters one is distinguished as a spectral parameter and is written as the argument $x$ of $\Rc(x)$. Denoting $R(x)=P\Rc(x)$, where $P$ is a permutation of two qubits, one constructs a staggered transfer matrix~\cite{vanicat18,ljubotina19} acting on $L$ qubits as
\begin{equation}
\!\!\!  T(x;\!u)\!\equiv \!{\rm tr}_\mathfrak{a}\!\left[\! R_{1,\mathfrak{a}}(x_+)\! R_{2,\mathfrak{a}}(x_-)\!R_{3,\mathfrak{a}}(x_+)\!\cdots\! R_{L,\mathfrak{a}}(x_-)\!\right]\!.
\end{equation}
Here, the first index $1,\ldots,L$ is a physical index that, as we will shortly see, corresponds to our qubits, while the second index $\mathfrak{a}$ denotes a single auxiliary qubit space over which we trace. $u\in\mathbbm{R}$ is a real inhomogeneity parameter, and $x_\pm\equiv x\pm \frac{u}{2}$. Due to the Yang-Baxter equation~\eqref{eq:YB}, such transfer matrices commute for different values of the spectral parameter $x \in\mathbbm{C}$. The brickwork circuit on $L$ qubits is then obtained as
\begin{equation}
\mathbb{U}=T(-\!\tfrac{u}{2};\!u)^{-1}T(\tfrac{u}{2};\!u)=\prod_{j=1}^{L/2}U_{2j,2j+1}\prod_{j=1}^{L/2}U_{2j-1,2j},
\end{equation}
where the indices are taken modulo $L$ in a circuit with periodic boundary conditions, and it turns out that the $U$ on the RHS is exactly equal to $\Rc$, i.e., $U=\Rc(u)$, and one can interpret $\mathbb{U}$ as a one-step propagator of a circuit~\cite{vanicat18}. Because $T(x;u)$ for different parameters $x$ commute (and thus also with $\mathbb{U}$), the conserved charges for a circuit can be obtained via the standard formalism of integrability. Namely, the logarithmic derivatives of the $L$-qubit transfer matrix $T(x;\!\tfrac{u}{2})$ at $x=\pm \tfrac{u}{2}$ yield two families of local conserved charges, which we denote by $Q_\ell^{\pm}(u)\equiv \partial_x^\ell \log T(x;\!u)|_{\pm\!\frac{u}{2}}$. These charges depend on the inhomogeneity parameter $u\in\mathbbm{R}$ and are invariant under translations for an even number of sites. For example, the first logarithmic derivatives give rise to two local charges
\begin{align}
\begin{aligned}
Q^{+}_1\!(u)&\!=\!\sum_{j=1}^{L/2}\! \partial_x\!\left[\!\Rc_{2j,2j\!+\!1}(x_+)\!\Rc_{2j\!+\!1,2j\!+\!2}(x_-)\!\right]_{\tfrac{u}{2}}\!U_{2j,2j\!+\!1}^{-1}\!,\\
Q^{-}_1\!(u)&\!=\!\sum_{j=1}^{L/2}\!U_{2j,2j\!+\!1}\!\partial_x\!\left[\!\Rc_{2j\!-\!1,2j}(x_+)\!\Rc_{2j,2j\!+\!1}(x_-)\!\right]_{-\!\tfrac{u}{2}}\!,
\end{aligned}
\label{eq:Q1}
\end{align}
whose local densities act on three adjacent qubits (see \app~\ref{sec:appB} for their explicit form). A similar property holds also for higher charges: the $\ell$-th conserved charge has local density supported on $r=2\ell+1$ neighboring sites. Together with the U(1) charge, i.e., magnetization, we find $r$ charges with support on at most $r$ lattice sites for any odd $r$. 

\section{From $\Uhaar$ to $\Rc$ matrix} 
We will show that $\Uhaar$ can be parameterized by the known unitary solution of the Yang-Baxter equation (\ref{eq:YB}) -- the $R$ matrix of the so-called asymmetric six-vertex model~\cite{sogo82,Marius13,vieira18},
\begin{equation}
\Rc(u)=\ee{\ii \beta u}\begin{pmatrix}
		1 & 0 & 0 & 0 \\
		0 & \ii b\,\ee{-\ii \xi u}  & -a\, \ee{-\ii \theta} & 0 \\
		0 & -a\, \ee{\ii \theta} &  \ii b\,\ee{\ii \xi u} & 0 \\
		0 & 0& 0 & 1
\end{pmatrix},
\label{eq:R}
\end{equation}
where $a$ and $b$ will be expressed in two different forms.
In what we call phase $I$, we will use
\begin{equation}
  \aI=\frac{\sin{u}}{\sin{(u+\ii \rho)}},\quad \bI=\frac{\sinh{\rho}}{\sin{(u+\ii \rho)}},
  \label{eq:abI}
\end{equation}
while in the phase $I\!I$ we have
\begin{equation}
  \aII=\frac{\sinh{u}}{\sinh{(u+\ii \rho)}},\quad \bII=\frac{\sin{\rho}}{\sinh{(u+\ii \rho)}}.
  \label{eq:abII}
\end{equation}
The reason to use two different forms is to always have a parametrization in terms of real parameters only. Alternatively, we could use just one form, say Eq.(\ref{eq:abI}), but with both purely real, as well as purely imaginary parameters. Our choice of $\Rc$, Eqs.~(\ref{eq:abI}) and~(\ref{eq:abII}), is therefore always parametrized by five real parameters $\beta,\xi,\theta,\rho$, and $u$, ensuring unitarity. Notably, it has one parameter more than the integrable Hamiltonian with local density given in Eq.~\eqref{eq:H}, since in the latter the staggered field $B$ cancels out due to translational invariance.
\begin{figure}[t!]
  \centerline{\includegraphics[width=3.0in]{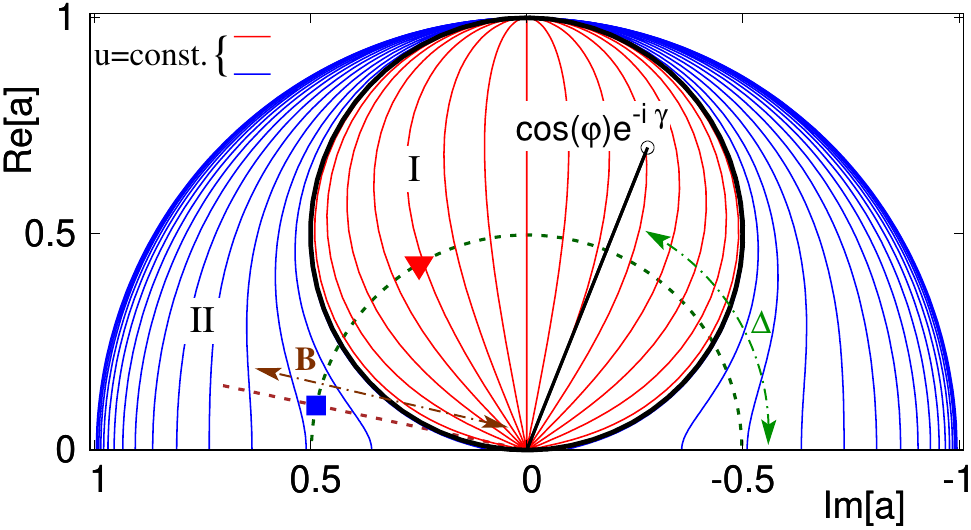}}
  \caption{Two dimensional projection of the phase diagram. Complex values of $a(u,\rho)$ must span the whole half-disc describing possible values of $\cos{\varphi}\ee{-\ii \gamma}$. Phase $I$ (in red) is separated from phase $I\!I$ (in blue) by the critical manifold (black circle). Curves are lines of constant $u$, the blue square and the red triangle are parameters shown in Fig.~\ref{fig:szm}. Green and brown dashed curves show how the circuit moves if one would vary the anisotropy $\Delta$ (green), or the staggered field $B$ (brown), fixing all other parameters.}
        \label{fig:polkrog}
\end{figure}

Using $\Uhaar= \Rc(u)$, we will explicitly express the above parameters in terms of Haar angles, thus showing that the $\Rc$ matrix spans all $\Uhaar$. First, we note that, in both phases, $a$ and $b$ satisfy $|a|^2+|b|^2=1$ and, since $u,\rho \in \mathbbm{R}$, also $a b^* \in \mathbbm{R}$. Therefore one can write $a=\cos{\tilde{\varphi}(u,\rho)}\ee{-\ii \gamma(u,\rho)}$ and $b=\sin{\tilde{\varphi}(u,\rho)}\ee{-\ii \gamma(u,\rho)}$. Comparing $\Rc$ and $\Uhaar$, we first deduce the relations
\begin{align}    
\beta u=&\delta,\quad  \theta=\vartheta,\quad \tilde{\varphi}=\varphi,\\
&\gamma=\delta-\alpha+\pi.   \label{eq:gamma}
\end{align}
Then, we must find real $u$ and $\rho$, such that
\begin{equation}
a=\cos{\varphi}\,\ee{-\ii \gamma},\qquad b=\sin\varphi\, \ee{-\ii\gamma}
\label{eq:a}
\end{equation}
are satisfied. There is a subtle point here: in order to be able to solve Eq.~\eqref{eq:a} for all possible angles $\delta$, $\alpha,\varphi$ with only either $\aI$ or $\aII$, we must use a freedom in the Haar block $\Vhaar$, Eq.~\eqref{eq:Haar}. Specifically, since $\Vhaar(\vartheta+\pi,\chi+\pi,\alpha+\pi,\varphi)=\Vhaar(\vartheta,\chi,\alpha,\varphi)$ holds, we can always bring $\gamma$, Eq.~\eqref{eq:gamma}, to the interval $\gamma \in [-\frac{\pi}{2},\frac{\pi}{2})$ (since $\gamma$ does not depend on $\vartheta$ or $\chi$, simply change them and $\alpha$ by an appropriate integer multiple of $\pi$). With this reduction, $\cos\varphi\, \ee{-\ii\gamma}$ in Eq.~\eqref{eq:a} fills the right half of the unit disk, which is exactly the set covered by $\aI$ and $\aII$ for real $u$ and $\rho$ (see Fig.~\ref{fig:polkrog}). This freedom in Haar block seems crucial in making the $R$ matrix span all $\Uhaar$. Without it, we would have to introduce an additional sign in the $R$ matrix~\cite{vieira18}. Using $\aI$ and $\bI$, Eq.~\eqref{eq:abI}, in Eq.~\eqref{eq:a}, we now obtain
\begin{equation}
  \uI=\arccos{\left(\frac{\sin{\gamma}}{\sin{\varphi}} \right)},\quad \rI={\rm arccosh}{\left(\frac{\cos{\gamma}}{\cos{\varphi}} \right)},
  \label{eq:uI}
\end{equation}
and finally, also $\xi u=\chi-\frac{\pi}{2}$. If, on the other hand, we take $\aII$ and $\bII$ from Eq.~\eqref{eq:abII}, we instead obtain
\begin{equation}
\uII\!={\rm arccosh}{\left(\frac{|\sin{\gamma}|}{\sin{\varphi}} \right)},\!\!\quad \rII\!=s \arccos{\!\left(\!\frac{\cos{\gamma}}{\cos{\varphi}} \right)},
  \label{eq:uII}
\end{equation}
where $s\!=\!{\rm sgn}(\sin{\gamma})$ is the sign of $\sin{\gamma}$, and $\xi u\!=\!\chi\!-\!s \frac{\pi}{2}$. Separation between the two solutions is as follows: if $\cos{\varphi}<\cos{\gamma}$ we are in the phase $I$, Eq.~\eqref{eq:uI}, while for $\cos{\varphi}>\cos{\gamma}$ we are in the phase $I\!I$, Eq.~\eqref{eq:uII}.

\section{The two phases}
Let us now discuss the physical properties of the two phases. In terms of parameters of the Hamiltonian~\eqref{eq:H}, the condition for the phase $I$ reads
\begin{equation}
\left| \frac{\sin{(2\tau\Delta)}}{\sin{(2\tau\sqrt{1+D^2+B^2})}}\sqrt{1+\frac{B^2}{1+D^2}}\right| > 1.
  \label{eq:cond}
\end{equation}
A two-parameter integrable XXZ circuit ($D=B=0$) of Ref.~\cite{ljubotina19} is recovered by setting $\beta=0$, $\xi=0$, and $\theta=\pi$. Specifically, the ballistic (``gapless'') phase of that circuit is contained in our phase $I\!I$, and the diffusive (``gapped'') one in the phase $I$. The integrable isotropic Heisenberg circuit, discussed in Ref.~\cite{vanicat18}, is contained in the critical manifold, and additionally requires substituting $u\to u\rho$ and taking the limit $\rho\to 0$.

Importantly, compared to the XXZ gates we have an integrable family with four parameters $\tau$, $\Delta$, $B$, $D$, and a three-dimensional critical manifold specified by the condition in Eq.~\eqref{eq:cond}. On the space of \MC gates the critical manifold therefore has a co-dimension one, which makes these circuits experimentally interesting: changing an arbitrary parameter, one will typically cross the critical manifold. One particular example is the transition when $\tau$ is varied (for the XXZ gate see Ref.~\cite{bertini23}): as seen in Eq.~\eqref{eq:cond}, upon changing $\tau$ (or, similarly, $\Delta$) one in fact repeatedly oscillates between the two phases, thus encountering an infinite number of such transitions. Such is the case also with $D$ -- it essentially just renormalizes the hopping to $\sqrt{1+D^2}$ (see \app~\ref{sec:appA}). On the other hand, changing $B$ is qualitatively different. Given parameters, there exists a point of no return $B_{\rm crit}$: further increasing $B$ beyond $B_{\rm crit}$ one cannot leave the phase $I$ anymore, approaching the origin $|a|=0$ (Fig.~\ref{fig:polkrog}) as $B \to \infty$ (the origin also corresponds to a continuum limit $\tau \to 0$ of the XXX gate~\cite{vanicat18}). Finally, we note that, if the unitary gate $\Uhaar$ is drawn according to the Haar measure, exactly half of the gates will correspond to phase $I$. If, on the other hand, one draws Hamiltonian parameters [cf. Eq.~\eqref{eq:H}], the ensemble will be, for nonzero $B$, biased towards phase $I$, due to the square-root in Eq.~\eqref{eq:cond}. 
 
\begin{figure}[t]
    \centerline{\includegraphics[width=3in]{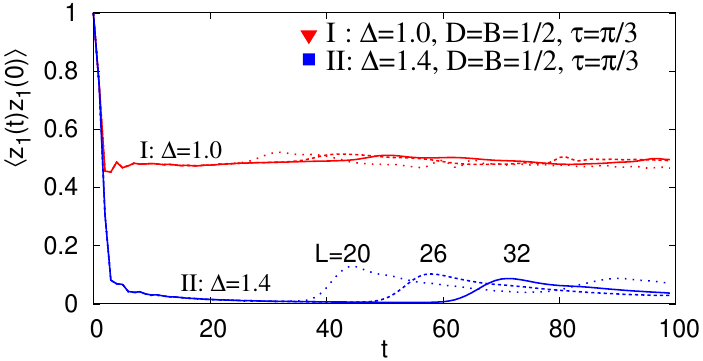}}
  \caption{Infinite-temperature autocorrelation function of a boundary magnetization in the two phases. Strong zero modes exist in phase $I$ (red), while correlations decay for the point in phase $I\!I$ (blue). Gate parameters correspond to the points shown in Fig.~\ref{fig:polkrog}.}
  \label{fig:szm}
\end{figure}

Since phases $I$ and $I\!I$ differ in the analytical structure of the $\Rc$ matrix, we expect also their physical properties to be different, similarly as for the XXZ gate~\cite{ljubotina19}. To illustrate different physics we touch upon the appearance of localized boundary modes that are robust to all magnetization-preserving gate perturbations, and which are present even at an infinite temperature. Their existence has been established in the gapped phase of the XXZ spin chain~\cite{fendley16}, and has been derived also in phase $I$ of the XXZ circuit~\cite{vernier24}. In Fig.~\ref{fig:szm} we numerically demonstrate that the infinite-temperature autocorrelation function of the boundary spin, $\tr{\sigma^z_1(t)\sigma^z_1}/2^L$, does not decay to zero in phase $I$ even in the presence of $D$ and $B$. Similarly, a fully polarized domain wall state, which is known to be ``frozen'' in time in the gapped XXZ chain~\cite{gochev}, exhibits similar behavior for generic MC gates in phase $I$. Lastly, in the Ruelle-Pollicott spectra of both phases, discussed below and shown in Fig.~\ref{fig:RP}, one can recognize the quasilocal conserved operators which, for instance, cause fractal ballistic transport in phase $I\!I$ of the XXZ gate. At present, the group structure required to analytically derive them for general parameters of the circuit is not known. The above observations support our conjecture that physical properties can differ between phases $I$ and $I\!I$.

\section{Ruelle-Pollicott spectrum} 

We now discuss the method of Ruelle-Pollicott (RP) resonances~\cite{Ruelle86,Pollicott85}, which is incidentally how we realized that \MC circuits are integrable. Introducing weak non-unitary perturbation to the propagator and analyzing its spectrum in the limit of zero perturbation, one may find frozen isolated eigenvalues inside the unit circle. Typically expected for chaotic systems, such Ruelle-Pollicott resonances signal exponential decay of correlation functions~\cite{gaspard,Braun}. Here, we apply the method to integrable systems, by writing the operator-evolution propagator in an infinite system, truncating it to $r$-local translationally invariant operators~\cite{Prosen,RP24}, and considering its RP spectrum (for an alternative non-unitarity by weak dissipation, less suitable for systems with conservation laws, see, e.g., recent Refs.~\cite{mori,sarang,curt}). More precisely, since brickwork circuits are two-site shift invariant, we can consider the sector with a given momentum $k$, spanned by operators $o_k[q]\equiv\sum_j \ee{\ii k j} {\cal S}^{2j} q$. Here, ${\cal S}$ is a one-site shift and $q$ is a local operator acting on at most $r$ consecutive sites. The matrix elements of the truncated propagator $\T$ between two basis elements $o_k[q]$ and $o_k[q']$ are then
\begin{equation}
\!\!\![\T]_{q',q}\!=\!\!\sum_{j\in\{-1,0,1\}}\!\! \ee{-\ii k j}\braket{{\cal S}^{2j} q'}{{\cal U}q},\quad\mathcal{U}q\!\equiv\! \mathbbm{U}^\dagger q \mathbbm{U},
\end{equation}
where $\braket{q'}{q}\equiv\tr{q'^\dagger q}/2^r$, while $q$ and $q'$ run over all $r$-local operators starting either on the even or odd sites (see Refs.~\cite{tobe,RP24} for details).
\begin{figure}
    \centerline{\includegraphics[width=1.6in]{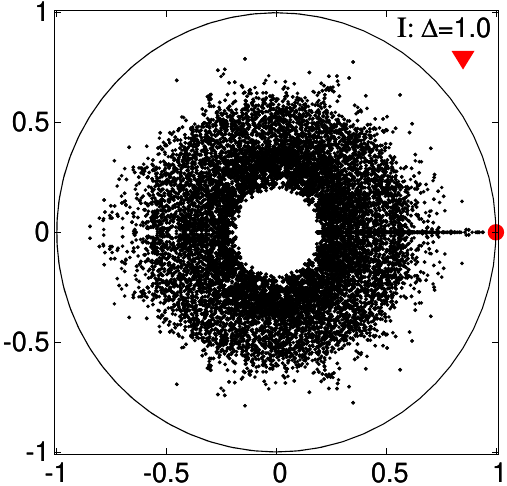}\includegraphics[width=1.6in]{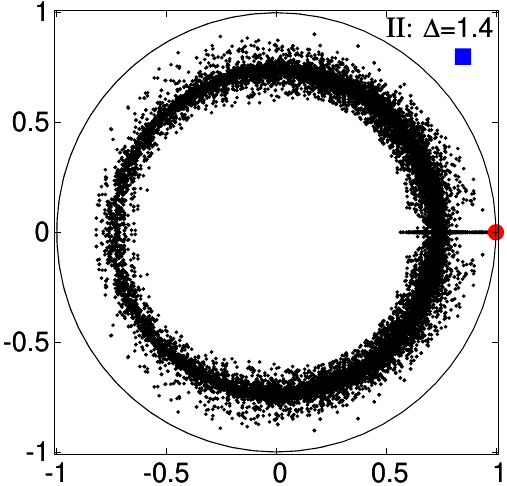}}
  \caption{RP spectrum for the same Hamiltonian parameters as shown in Fig.~\ref{fig:szm}. The red point marks $\lambda=1$ (7 times degenerate for the shown $r=7$).}
  \label{fig:RP}
\end{figure}
A representative spectrum of $\T$ is shown in Fig.~\ref{fig:RP} for $k=0$. We find that the eigenvalue $1$ is $r$-times degenerate (for odd $r$), the corresponding eigenvectors being the magnetization and the local conserved charges $Q_\ell^\pm(u)$ with $2\ell+1\le r$ [cf. Eq.~\eqref{eq:Q1}]. Another prominent feature are the eigenvalues which, when $r$ is increased, get more numerous and exponentially closer in modulus to $1$. The corresponding eigenvectors are quasilocal conserved operators: for a finite $r$ they are almost conserved, the violation coming from their exponential tails. For further details see \app~\ref{sec:appC}.

\section{Conclusion} 

We have shown that a two-parameter family of U(1) invariant integrable circuits, for example with the XXZ gate, can be extended to a four-parameter (excluding the field $M$) integrable family containing all brickwork circuits with magnetization-preserving two-qubit gates. We find this surprising, since integrability is associated with fine-tuning, while chaotic dynamics is considered generic. We conjecture that the integrable family described herein is the largest possible among the homogeneous qubit circuits, and is furthermore larger than the largest integrable spin chain class. It could play a similar role in quantum computation as other integrable many-body systems do in their respective domains.

Many new questions are opened, first and foremost, what are the properties of these models? For instance, for the two-parameter subset describing the integrable XXZ circuit~\cite{ljubotina19} we know that, crossing from one phase to another, spin transport properties in the zero-magnetization sector change drastically. In our extended family we have two new parameters, so one can perhaps expect even richer behavior, e.g., the staggered field $B$ seems to have a qualitatively different effect than the other three parameters. The chiral Dzyaloshinskii-Moriya term could perhaps lead to similar effects of broken space-reflection symmetry as in recently studied circuits and Hamiltonian systems in Refs.~\cite{ratchet,vedika20,richelli24}. 

The critical manifold is also of particular interest. It contains the known~\cite{vanicat18} one-parameter XXX gate for which Kardar-Parisi-Zhang~\cite{KPZ} equilibrium fluctuations of the two-point correlations have been observed~\cite{kpz} due to integrability and SU(2) symmetry~\cite{enej}. We expect that similar physics can be observed also at non-isotropic values of the parameters ($\Delta \neq 1$).

Yet another open direction are integrable nonunitary circuits~\cite{sa21,gritsev24,paletta24} from the same $R$ matrix, but with complex parameters. Here one should consider Sklyanin's boundary reflection equations~\cite{sklyanin88}, which put constraints on integrable systems with open boundary conditions. Concerning open boundary conditions, the existence of strong zero modes also calls for further studies.

Experimental possibilities are equally exciting. While in the past the XXZ- or XXX-type gates have been targeted~\cite{john23,maruyoshi23,google24} due to their interesting physics, our results show that any magnetization-conserving gate will do. No fine-tuning is therefore required, and what is more, to be at a critical point one just has to vary any generic parameter, like the gate duration or the anisotropy.

Having a large integrable family of circuits also poses a question regarding robustness of their properties to integrability breaking. Could a slow thermalization in an apparently chaotic model with different \MC gates, recently observed in Ref.~\cite{cheryne23}, be related to being close (on average) to the integrable class of circuits discussed here? 

We also show the usefulness of the truncated operator propagator -- an object usually used in studies of Ruelle-Pollicott resonances of chaotic systems. Compared to the level-spacing statistics, which can be plagued by hard-to-identify symmetries and other issues~\cite{not}, it directly targets arguably the most important object in integrable systems -- the conserved operators -- and could be used to identify other new integrable systems.

Lastly, we point at an interesting observation discussed in \app~\ref{sec:appA}, which is relevant for the level spacing statistics in Fig.~\ref{fig:lsd}. Namely, when boundary conditions in a generic non-integrable (even disordered) brickwall circuit with \MC gates are changed from open to periodic, an unconventional time-reversal symmetry is broken, changing the spectral statistics from the orthogonal to the unitary symmetry class.

We thank Chiara Paletta, Enej Ilievski, and Toma\v{z} Prosen for useful comments on the manuscript. We acknowledge support by Grants No.~J1-4385 and No.~P1-0402 from Slovenian Research Agency. L.Z. also acknowledges support by ERC Advanced Grant No. 101096208--QUEST.

\clearpage
\appendix

\PRL

\setcounter{secnumdepth}{2}
\renewcommand{\thesubsection}{\Alph{section}.\arabic{subsection}}

\section{Time-reversal symmetry breaking by periodic boundary conditions}
\label{sec:appA}

Let us show that a circuit with brickwall configuration of (in general) different \MC gates and open boundary conditions has an anti-unitary time-reversal symmetry.

We start by considering a single \MC gate $U$. It is easy to see that any MC gate has an anti-unitary time-reversal symmetry $\mathcal T_1 U \mathcal T_1^{-1} = U^\dagger$ of the form
\begin{equation}
  \mathcal T_1 = W_\vartheta K,\quad W_\vartheta=\exp\left[- \ii \tfrac{\vartheta}{2} \left(\sigma_2^z - \sigma_1^z\right)\right],
\end{equation}
where $\vartheta$ is the off-diagonal angle parameter of the gate in Eq.~\eqref{eq:Haar}, and $K$ is a complex conjugation in the computational basis. Analogously, we can also bring the generator $h$ (\ref{eq:H}) to a manifestly real form by rotating out the Dzyaloshinskii-Moriya term $D$. Namely, choosing $\vartheta$ such that it satisfies $\tan{(2\vartheta)}=-D$, we have
\begin{equation}
 \!\!\! W_\vartheta h_{1,2}(J=1,\Delta,B,D) W^\dagger_\vartheta\! =\! h_{1,2}(\sqrt{1\!+\!D^2},\Delta,B,0),
\end{equation}
where, for clarity, we reinstated the hopping $J$ -- a prefactor in front of $\sx_1\sx_2+\sy_1\sy_2$ in $h_{1,2}$ (we use $J=1$ throughout the paper). A suitable rotation therefore removes gate's $D$, thus renormalizing the hopping.

\begin{figure}[t!]
	\centerline{\includegraphics[width=0.427\textwidth]{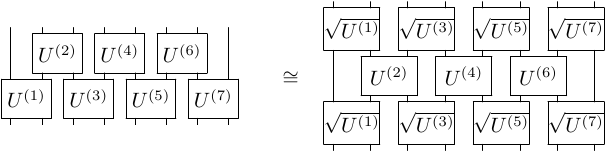}}
  \caption{A brickwall circuit with open boundary conditions (left) has the same spectrum as a 3-layer brickwall circuit (right), where the 1st and 3rd layers consist of square
roots of gates from the 1st brickwall layer.}
	\label{fig:time_reversal}
\end{figure}

Rotation by local $W_\vartheta$ can be extended to a brickwall circuit with open boundaries (see Fig.~\ref{fig:time_reversal} on the left)
\begin{equation}
	\mathbb U = \prod_{j = 1}^{L/2-1} U^{(2j)}_{2j, 2j + 1} \prod_{j = 1}^{L/2} U^{(2j - 1)}_{2j - 1, 2j},
\end{equation}
where all $U^{(j)}$ are in general different \MC gates. We define
\begin{equation}
  \mathcal T = \prod_{j = 1}^L \exp\left(-\ii \sum_{k = 1}^{j - 1}\vartheta^{(k)} \sigma_j^z\right) K, \label{eq:time_rev}
\end{equation}
where $\vartheta^{(j)}$ is the parameter of the gate $U^{(j)}$. Crucial property of the above rotation is that the difference of rotation angles on qubits $j$ and $j+1$ is precisely $\vartheta^{(j)}$, i.e., the angle of the gate acting on those two qubits. By acting with $\mathcal T$ on $\mathbb U$, we map each local gate $U^{(j)}$ to its Hermitian
adjoint $U^{(j)\dagger}$, but do not change the order of layers.
This means that $\mathcal T$ \textit{is not} yet a time-reversal
symmetry of $\mathbb U$, i.e., $\mathcal T \mathbb U \mathcal T^{-1}\neq \mathbb{U}^\dagger$.
However, we can construct a different circuit $\tilde{\mathbb U}$, whose
spectrum is the same as the spectrum of $\mathbb U$, and for which this $\mathcal T$
is a time-reversal symmetry. We can construct such a circuit by recalling that
cyclic permutations do not change the spectrum, i.e., $\sigma(ABC) =
\sigma(BCA)$~\cite{spacetime, jas21}, where $\sigma(\bullet)$ denotes the
spectrum and $A, B, C$ are arbitrary square matrices. We shall denote such spectral equivalence by $ABC
\cong BCA$. We can now factor the first layer of $\mathbb U$ into two layers,
consisting of some square roots of local gates $\sqrt{U^{(j)}}$ and then
cyclically permute the layers to obtain the circuit $\tilde{\mathbb U}$ (see Fig.~\ref{fig:time_reversal}),
\begin{equation}
  \tilde{\mathbb U}\!=\!\prod_{j = 1}^{L/2} \sqrt{U^{(2j\!-\!1)}_{2j\!-\!1, 2j}}\! \prod_{j\!=\!1}^{L/2\!-\!1} U^{(2j)}_{2j, 2j\!+\!1}\! \prod_{j\!=\!1}^{L/2} \sqrt{U^{(2j\!-\!1)}_{2j\!-\!1, 2j}} \!\cong\! \mathbb U. \label{eq:equiv_circ}
\end{equation}
It is now easy to see that 
\begin{equation}
  \mathcal T \tilde{\mathbb U} \mathcal T^{-1} = \tilde{\mathbb U}^\dagger,
\end{equation}
since the order of application of gates (``geometry'' of the circuit) is the
same for $\tilde{\mathbb U}$ and $\tilde{\mathbb U}^\dagger$. It is important
to note that $\sqrt{U^{(j)}}$ can be chosen in a way that they are MC, e.g., by
taking $\sqrt{U} = \exp(-\ii \frac{\tau}{2} h)$ in the Hamiltonian
parametrization $U = \exp(-\ii \tau h)$ from Eq.~\eqref{eq:H}.

Our brickwall circuit (right panel of Fig.~\ref{fig:lsd}) is a special case in which the gates in each layer are the same. With open boundary conditions it has the same spectrum as the circuit with a time-reversal symmetry. Therefore, we expect that all functions of the spectrum
will display behavior consistent with a time-reversal symmetry. Since $\mathcal
T^2 = 1$, we observe the level-spacing distribution consistent with the
circular orthogonal ensemble (COE)~\cite{haake} in Fig.~\ref{fig:lsd} (red curve ``COE'' is the Wigner's surmise). Our
construction works in arbitrary geometry with open
boundary conditions since all such geometries have the same
spectra~\cite{jas21}. In geometries with periodic boundary conditions the
construction in Eq.~\eqref{eq:time_rev} does not work in general since we do
not have an additional free parameter to correctly map the gate acting on spins
$L$ and $1$, $U^{(L)}_{L, 1}$. The construction would work only in special cases, for instance, for the brickwall
geometry in the fine-tuned case where $\vartheta^{(L)} = -\sum_{j = 1}^{L - 1}
\vartheta^{(j)}$ (modulo $2 \pi$).

Brickwall circuits with \MC gates therefore have a peculiar property affecting their spectra, and thereby also often-used indicators of quantum chaos~\cite{haake} (e.g., the level-spacing distribution or the spectral form factor): with open boundaries they are in the orthogonal symmetry class, with periodic in the unitary class (except for special cases). Bulk physics, of course, does not depend on boundary conditions.

\section{Local densities of $Q^{\pm}_1(u)$}
\label{sec:appB}

We report the explicit form of the local charges  in Eq.~\eqref{eq:Q1}. Denote $h_{j,j\!+\!1}^{\rm XX}=\sx_j\sx_{j\!+\!1}\!+\!\sy_j\sy_{j\!+\!1}$ and $h_{j,j\!+\!1}^{\rm DM}=\sx_j\sy_{j\!+\!1}\!-\!\sy_j\sx_{j\!+\!1}$. In the phase $I$ we now have $Q_1^+(u)=\sum_j q_{2j\!+\!1}^{(1;+)}(u)$ and $Q_1^-(u)=\sum_j q_{2j}^{(1;-)}(u)$, where
\begin{widetext}
\begin{align}
q^{(1;\pm)}_j(u)\!=&\frac{1}{2\ii (\cos 2u\!-\!\cosh2\rho)}\Big\{2\cos u\sinh\rho \big[\cos(\theta\!\pm\!\xi u)h_{j\!-\!1,j}^{\rm XX}+\cos(\theta\!\mp\!\xi u)h_{j,j\!+\!1}^{\rm XX}
-\sin(\theta\!\pm\!\xi u)h^{\rm DM}_{j\!-\!1,j}-\sin(\theta\!\mp\!\xi u)h^{\rm DM}_{j,j\!+\!1} \notag\\
&-\frac{\cosh\rho}{\cos u}(\sz_{j\!-\!1}\sz_{j}+\sz_{j}\sz_{j\!+\!1})
\big]
- 2\sin^2 u\coth\rho\big[\cos(2\theta) h_{j\!-\!1,j\!+\!1}^{\rm XX}-\sin(2\theta) h_{j\!-\!1,j\!+\!1}^{\rm DM}+\sz_{j\!-\!1}\sz_{j\!+\!1}\big]\notag\\
&\mp 2 \sin u\cosh\rho \big[\sin(\theta\!\pm\!\xi u)h^{\rm XX}_{j\!-\!1,j}+\cos(\theta\!\pm\!\xi u)h_{j\!-\!1,j}^{\rm DM}\big]\sz_{j\!+\!1}\mp\sin2u\big[\sin(2\theta)h_{j\!-\!1,j\!+\!1}^{\rm XX}+\cos(2\theta)h_{j\!-\!1,j\!+\!1}^{\rm DM}\big]\sz_{j}\notag\\
&\mp 2\sin u\cosh\rho \big[\sin(\theta\!\mp\!\xi u)h_{j,j\!+\!1}^{\rm XX}+\cos(\theta\!\mp\!\xi u) h_{j,j\!+\!1}^{\rm DM}
\big]\sz_{j\!-\!1}
\Big\}.
\end{align}
The charges in the phase $I\!I$ are obtained by substituting $u\to \ii u$, $\rho\to \ii \rho$, $\xi \to -\ii \xi$, and $\beta \to -\ii \beta$.
\end{widetext}

\section{Ruelle-Pollicott spectrum}
\label{sec:appC}

\begin{figure}
    \centerline{\includegraphics[width=2.4in]{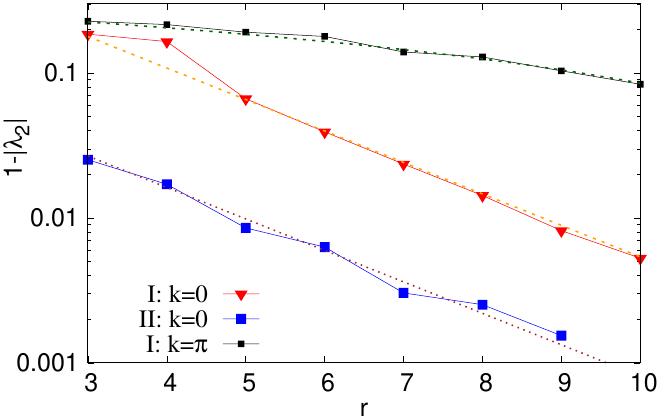}}
  \caption{Gaps of $\T$ for different operator support $r$. Gate parameters are the same as in Fig.~\ref{fig:RP}. Two straight lines fitting zero-momentum gaps are exponentials, green dashed curve approximating the data for $k=\pi$ is a linear function.}
  \label{fig:gaps}
\end{figure}

In addition to the discussion in the main text, few other observations about the spectrum of the truncated propagator are in order. As already mentioned, in the sector with $k=0$ we observe a number of eigenvalues that are exponentially close in modulus to $1$. In Fig.~\ref{fig:gaps} we show the gap, i.e., $1-|\lambda_2|$, of the one closest to $1$, denoted here simply as $\lambda_2$ (remember that $\lambda_1=1$ is degenerate).
We can see that, in the $k=0$ sector, $\lambda_2$ goes as $1-\lambda_2 \approx c\ee{-0.5 r}$ with $c \approx 0.8$ for the shown point in the phase $I$ (red triangles), and $c \approx 0.12$ for the point in the phase $I\!I$ (blue squares). Different prefactors suggest that there are more quasilocal conserved operators below a fixed localization length in the ``ballistic'' phase II. 

\begin{figure}[b!]
    \centerline{\includegraphics[width=1.45in]{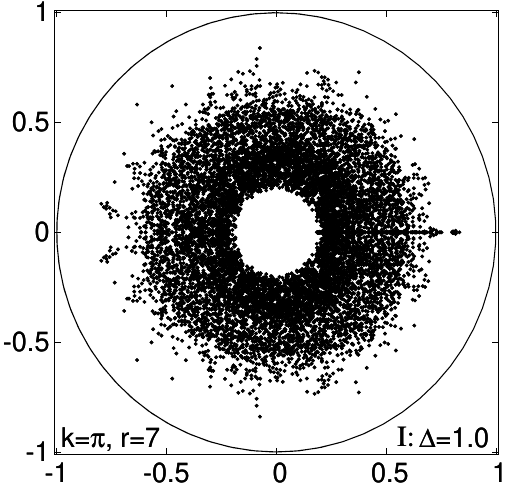}\includegraphics[width=1.45in]{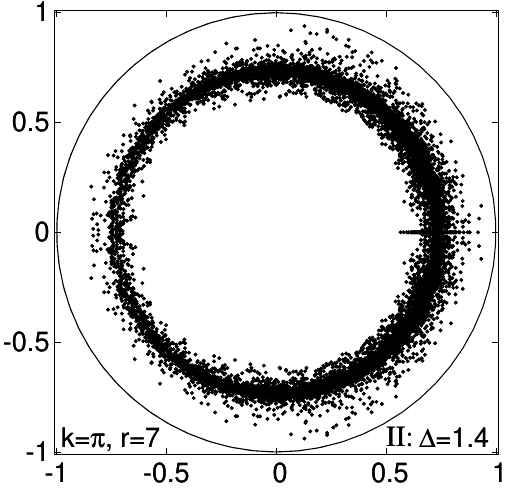}}
  \caption{RP spectrum in the sector with momentum $k=\pi$ for the same Hamiltonian parameters as in Fig.~\ref{fig:RP}, i.e., $\tau=\frac{\pi}{3}$ and $B=D=\frac{1}{2}$, and $\Delta=1.0$ (left, red triangle in Fig.~\ref{fig:polkrog}), or $\Delta=1.4$ (right, blue square in Fig.~\ref{fig:polkrog}).}
  \label{fig:RPk1}
\end{figure}
\begin{figure}[b!]
    \centerline{\includegraphics[width=2.4in]{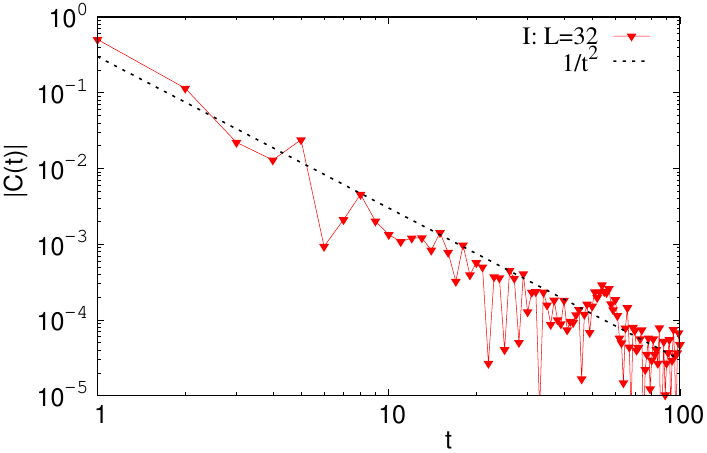}}
  \caption{Correlation function of the staggered magnetization $S$ decays as a power law. Gate parameters are the same as for the phase $I$ point in Fig.~\ref{fig:RP}.}
  \label{fig:power}
\end{figure}

In the nonzero momentum sector the spectrum of $\T$ looks similar to the one in $k=0$, except that there are no eigenvalues corresponding to quasilocal conserved operators. An example of spectra for $k=\pi$ is shown in Fig.~\ref{fig:RPk1}. Despite the absence of eigenvalues very close to $1$, as one increases $r$ the largest $|\lambda_1|$ again approaches $1$, however, this time only linearly in $r$ (see Fig.~\ref{fig:gaps} for $k=\pi$ and parameters from the phase $I$). This is in line with integrability and expectation that one does not have any exponential relaxation of correlation functions. We have checked numerically that the infinite-temperature autocorrelation function $C(t)=\ave{S(t)S(0)}/L$ of staggered magnetization $S$ from the sector $k=\pi$, $S\!\equiv\! \sum_{j=1}^{L/2}(\!-\!1)^j(\sz_{2j\!-\!1}\!+\!\sz_{2j})$,
decays as a power law, see Fig.~\ref{fig:power}. Visible oscillations are not noise (finite-size fluctuations are of the order $\sim 10^{-5}$), but rather oscillations due to complex pairs (or quadruples) of eigenvalues of $\mathbb T(k = \pi)$ with the same modulus.


\begin{thebibliography}{99}
\bibitem{not} Note that the level-spacing distribution (LSD) should not be used as a definition: one can have chaotic LSD in integrable systems~\cite{seligman}, or have integrable systems without Poisson statistics~\cite{berry}, or Poisson statistics in chaotic circuits with unresolved space-time symmetries~\cite{spacetime}.
\bibitem{foot0} All matrices are written in a two qubit basis ordered as $\{ \ket{0_10_2},\ket{1_10_2},\ket{0_11_2},\ket{1_11_2}\}$.
\bibitem{foot5} Note that in such Hamiltonians there are only three nontrivial parameters: the staggered field $B$ cancels out due to invariance under one-site translations.
\end{thebibliography}
\end{document}